\documentstyle[12pt]{article}
\def\be{\begin{equation}}
\def\ee{\end{equation}}
\begin{document}
\baselineskip=22pt plus 1pt minus 1pt

\begin{flushright}
{\tt FTUV/93-53\\
     IFIC/93-34\\
     December 1993}
\end{flushright}

\begin{center}
{ \Large \bf Discretization of the phase space for a $q$-deformed harmonic
 oscillator with $q$ a root of unity}\\
\bigskip\bigskip\bigskip

{Dennis Bonatsos $^*$ \footnote[1]{Permanent address: Institute of Nuclear
Physics, N.C.S.R. ``Demokritos'', GR-15310 Aghia Paraskevi, Attiki, Greece},
 C. Daskaloyannis $^+$, Demosthenes Ellinas $^{\#}$ and Amand Faessler $^*$ }

\bigskip
{$^*$ Institut f\"ur  Theoretische  Physik, Universit\"at T\"ubingen,
Auf der Morgenstelle 14, D--72076 T\"ubingen, F. R. Germany}

{$^+$ Department of Physics, Aristotle University of
Thessaloniki, GR--54006 Thessaloniki, Greece }

{$^{\#}$ Departamento de Fisica Te\'orica and IFIC, Centro Mixto Universidad
de Valencia -- CSIC, E--46100 Burjasot, Valencia, Spain}
\bigskip

\end{center}

\bigskip\bigskip

\begin{center}
{\bf Abstract}
\end{center}

The ``position'' and ``momentum'' operators for the q-deformed oscillator
with q being a root of unity are proved to have  discrete eigenvalues
which are roots of deformed Hermite polynomials. The Fourier transform
connecting the ``position'' and ``momentum'' representations is also
found.  The phase
space of this oscillator has a lattice structure, which is a
non-uniformly distributed grid. Non-equidistant  lattice structures
 also occur in the cases of the truncated harmonic oscillator and of the
q-deformed parafermionic oscillator, while the parafermionic
oscillator corresponds to a uniformly distributed grid.

\vfill\eject

Quantum groups and algebras \cite{Dri},
which from the mathematical point of view are Hopf algebras, are recently
attracting much attention in physics, and especially after its introduction
 the so-called q-deformed harmonic oscillator \cite{Mac,Bie}, has been
the subject of intensive study.
The introduction of the q-deformed oscillator  has posed the
 question of the construction of a $q$-deformed ``coordinate'' or
``momentum'' representation, in analogy to the ordinary quantum mechanics.
This problem was initially studied in the context of the
$Q$-deformed oscillator, introduced earlier by Arik and Coon \cite{AC}
(see also \cite{Mac}). The properties of the $a^+ +a$ operator (which is
proportional to the position operator in usual quantum mechanics) for
the q-deformed oscillator with $q$ real have been considered in \cite{Kli}.
The case of the $q$-deformed oscillator with $q$ being a root of unity
\cite{Hayashi,FT} is,
however,  qualitatively different from the case of $q$ real or the case of
$Q$-deformation \cite{AC}, because in the first case the Hilbert space is
finite dimensional, while in the latter the carrier Fock space is an
infinite dimensional Hilbert space.

In the present letter we consider the q-deformed oscillator \cite{Mac,Bie}
for $q$ being a root of unity.
 In this case we show that the finite
dimensionality of the Fock space induces a discretization of the spectrum of
 the ``position'' and ``momentum'' operators, the phase space being replaced
 by a grid (a lattice) with non-equidistant points which are identified as the
nodes of certain deformed Hermite polynomials. Similar non-uniformly
spaced lattices are shown to occur also in the case
of the truncated harmonic oscillator \cite{Fig}, i.e. the harmonic oscillator
 with truncated finite dimensional spectrum, and the q-deformed parafermionic
oscillator {\cite{FV}, while the usual parafermionic oscillator \cite{OK}
corresponds to
a lattice structure of the phase space with equidistant points.

The q-deformed harmonic oscillator is determined by the creation and
annihilation operators $a^+$ and $a$, which satisfy the relations
\be
 a a^+ - q^{\mp 1/2} a^+ a = q^{\pm N/2}, \qquad [N,a^+]=a^+,
\qquad [N,a]=-a,
\ee
where $N$ is the number operator and q-numbers are defined as
\be
[x] = {q^{x/2}-q^{-x/2}\over q^{1/2}-q^{-1/2}}.
\label{eq:q-str}
\ee
The relevant Fock space representation is determined by the eigenvectors
$$
a | 0>=0, \qquad |n> = {(a^+)^n\over ([n]!)^{1/2}} |0>,
$$
where the q-factorial is defined as
$$
[n]!= [n] [n-1] \ldots [1].
$$
The action of the operators on the basis vectors is given by
\be
 a^+ |n>= \sqrt{[n+1]} |n+1> , \qquad a |n> = \sqrt{[n]} |n-1>, \qquad
N |n> = n |n>.
\label{eq:q-osc2}
\ee
One can easily show that  for this representation
$$
 a^+ a = [N], \qquad a a^+ = [N+1].
$$
The Hamiltonian of the q-deformed oscillator is
$$
H = {1\over 2} (a a^+ + a^+ a),
$$
and its eigenvalues are
$$
E(n)= {1\over 2} ([n]+[n+1]).
$$

Let us consider the case in which $q$ is a root of unity, i.e.
$$
q=\exp({2\pi i\over n_d}),
$$
where $n_d$ is a natural number and $n_d\geq 2$. q-numbers then obtain
the form
\be
 [x] = {\sin({\pi x\over n_d})\over \sin({\pi \over n_d})}.
\label{eq:sf-q}
\ee
It is clear in this case that $q^{n_d}=1$ and  $[n_d]=0$, so that
in addition one has $[n_d \quad k] =0$ and $[n_d+k] =[k]$, where $k$ integer.
This implies \cite{Ellinas} that the Fock space matrix representation of the
generators as
given above, is reducible into a
block-diagonal form  with matrices along the diagonal
$$ a= \sum _{m=1}^{n_d-1} \sqrt{[m]} \vert m-1> < m\vert, \qquad
  a^+ = \sum _{m=0}^{n_d-2} \sqrt{[m+1]} \vert m+1> < m\vert , $$
$$ \qquad N= \sum _{m=0}^{n_d-1} m \vert m> <m\vert .$$
In other words, the Fock space is limited to $|n>$ with $n=0$, 1, 2,
\dots $n_d-1$. The nilpotency of $a$ and $a^+$ ($a^{n_d} = (a^+)^{n_d} =0$)
implies that the q-deformed oscillator with $q$ a $n_d$th root of unity
interpolates between the Pauli matrices ($n_d=2$) \cite{fermionic} and the
usual harmonic oscillator ($n_d \rightarrow \infty$).

In analogy to the case of the usual harmonic oscillator, let us consider
the ``position'' operator (see also \cite{Kli})
\be
X = {a+a^+\over \sqrt{2}}.
\label{eq:X-def}
\ee
This choice being quite arbitrary, we will show later in this letter
(starting from eq. (\ref{eq:genX}) that the basic features in the
subsequent discussion do not depend on this specific choice, which leads
to finite lattice phase spaces with open ends. A different possibility
is provided by the so called ``polar decomposition'' (see \cite{Ellinas,Ell}
and references therein), which is unique and leads to closed finite lattice
phase spaces, i.e. to a latticed torus.

Let then $|x>$ and $x$ be the eigenvectors and the corresponding eigenvalues
of the operator $X$, satisfying
\be
 X |x> = x|x>.
\label{eq:X}
\ee
It is clear that, when $q$ is a root of unity,
$|x>$ are vectors in the, as mentioned above, $n_d$-dimensional Fock space.
Therefore their general form is given by
\be
|x> = \sum_{n=0}^{n_d-1} {c_n\over \sqrt{[n]!}} |n> =
         \sum_{n=0}^{n_d-1} c_n {(a^+)^n\over [n]!} |0> .
\label{eq:x_exp}
\ee
We are now going to determine recursion relations for the coefficients
$c_n$ by using the Lanczos algorithm (\cite{Lan,Hay}, see \cite{FLTV}
for a group theoretical formulation of it).
Acting with the operator $X$ of eq. (\ref{eq:X-def})
 on the basis vectors of eq. (\ref{eq:x_exp})
and making use of eq. (\ref{eq:q-osc2}) one finds
\be
X|x> = {1\over \sqrt{2}} \sum_{n=0}^{n_d-2} {c_{n+1}\over \sqrt{[n]!}} |n>
  + {1\over \sqrt{2}} \sum_{n=1}^{n_d-1} {c_{n-1}\over \sqrt{[n]!}} [n] |n>.
\label{eq:Xx}
\ee
{}From eqs (\ref{eq:X}) and (\ref{eq:x_exp}) this should be equal to
\be
 x|x> = \sum_{n=0}^{n_d-1} {x c_n\over \sqrt{[n]!}} |n>.
\label{eq:xx}
\ee
Equating the rhs of eqs (\ref{eq:Xx}) and (\ref{eq:xx})
 one obtains the following
recurrence relations for the coefficients $c_i$

$$ c_1 =           \sqrt{2} x c_0,$$
\be
 c_{n+1} =   \sqrt{2} x c_{n}- [n] c_{n-1},
\label{eq:cn}
\ee
\be
c_{n_d}=0.
\label{eq:cnd}
\ee
These  recurrence relations imply that each $c_n$ is a polynomial
of order $n$ depending on $x$. The condition of eq. (\ref{eq:cnd})
is satisfied only by a set of discrete  values of the
eigenvalue $x$, which are the roots of the $n_d$--order polynomial
$c_{n_d}=c_{n_d}(x)$. The number of these roots is equal to the
order $n_d$ of the corresponding polynomial and these roots are
real ones because the operator $X$ defined by eq. (\ref{eq:X-def}) is
a Hermitian operator.
 This shows that $q$ being a root of unity induces a
 discretization of the spectrum of the ``position'' operator $X$.
The discrete values of the position $x$ are indexed by
the following convention:
$$
x_m, \quad m=-J,-J+1,\ldots,J-1,J \quad
\mbox{where}\quad
n_d=2 J+1,
$$
and
$$ c_{n_d}(x_m)=0. $$
The equations obtained for the first few values of $n$ are listed
below:
$$ c_{2}= \left( 2 x^2-[1] \right) c_0, $$
$$ c_{3}=\left( \left(\sqrt{2} x\right)^3 - ([1]+[2])\sqrt{2}x \right)c_0, $$
$$ c_{4}=\left( 4 x^4 - 2([1]+[2]+[3]) x^2 + [1][3]\right)c_0,$$
$$ c_{5}=\left( \left(\sqrt{2} x\right)^5
-([1]+[2]+[3]+[4]) \left(\sqrt{2}x\right)^3 +([1][3]+[1][4]+[2][4])
 \sqrt{2}x\right)c_0,$$
$$
\begin{array}{l}
 c_{6}=\Big(  8 x^6
-4([1]+[2]+[3]+[4]+[5]) x^4+\\
 +2 ([1][3]+[1][4]+[1][5]+[2][4]
      +[2][5]+[3][5]) x^2 -[1][3][5]\Big)c_0.
\end{array}$$
It is worth remarking at this point that the polynomials appearing in
the above equations are some kind of q-deformed Hermite polynomials
up to some normalization factor,
occurring from
the recursion relation
\be
H_{n+1}(x) = 2 x H_n(x) -2[n] H_{n-1}(x),
\label{eq:Hermite}
\ee
with $H_{-1}(x)=0$, $H_0(x)=1$.
 In fact this recursion relation induces the relation between
the Hermite polynomials and the functions $c_n$:
\be
c_n(x_m)=
\frac{2^{-n/2}}{ \sqrt{\cal N}_m } H_n(x_m),
\label{eq:relation}
\ee
where ${\cal N}_m$ is a normalization factor. Furthermore, one has
$$
H_{n_d}(x_m)=0, \quad  m=-J,-J+1,\ldots , J-1, J,
$$
i.e. the eigenvalues of the ``position'' operator $X$ in a $n_d$-dimensional
space are the roots of the corresponding deformed Hermite polynomial
$H_{n_d}$.

It is also worth noticing that the recursion relation of
eq. (\ref{eq:Hermite}) is similar
(but not identical) to the one obtained in ref.
\cite{Jeugt}. (We are going to discuss this correspondence again near the
end of this paper.) The q-Hermite polynomials
of \cite{Jeugt}  are also related to the ones introduced in refs.
 \cite{Allaway}, \cite{Bressoud}.
Other forms of q-Hermite polynomials have been introduced in refs.
\cite{Exton,Srivastava,AlSalam,Atakishiev,FV2,Minahan,Chang,DK}.

The Hermite polynomials defined by the recurrence relation of eq.
(\ref{eq:Hermite}) are odd (or even) functions of $x$ for n odd (or even),
therefore their eigenvalues are symmetric around the origin:
$$
x_{-m}= - x_m.
$$
{}From the recurrence relation the Darboux-Christofell \cite{Sze} equation can
be
proved:
\be
\begin{array}{l}
\rho(x,y)=\displaystyle \sum\limits_{n=0}^{n_d-1}
\frac{H_n(x)H_n(y)}{[n]!2^n} =\\
\displaystyle \frac{1}{[n_d-1]!2^{n_d-1}} \cdot
\frac{H_{n_d}(x)H_{n_d-1}(y)-H_{n_d-1}(x)H_{n_d}(y)}{x-y}.
\end{array}
\label{eq:Darboux}
\ee
The Darboux-Christofell equation implies:
$$
\rho(x,x_m)=\displaystyle \sum\limits_{n=0}^{n_d-1}
\frac{H_n(x)H_n(x_m)}{[n]!2^n} =
\displaystyle \frac{1}{[n_d-1]!2^{n_d-1}} \cdot
\frac{H_{n_d}(x)H_{n_d-1}(x_m)}{x-x_m}.
$$
If $x\to x_m$ then:
$$
\rho(x_m,x_m)=\displaystyle \sum\limits_{n=0}^{n_d-1}
\frac{H^2_n(x_m)}{[n]!2^n} =
\displaystyle \frac{1}{[n_d-1]!2^{n_d-1}} \cdot
{H'_{n_d}(x_m)H_{n_d-1}(x_m)}={\cal N}_m.
$$
The above equation means that $H'_{n_d}(x_m)\ne 0$ and therefore
the roots $x_m$ of the Hermite polynomials are simple ones.

The normalized coordinate eigenvector is then given by:
\be
\left\vert  x_m \right>
=
{1 \over \sqrt{{\cal N}_m }}
\sum\limits_{n=0}^{n_d-1}
{ {H_n \left(  x_m \right)} \over
\sqrt{\left[ n \right]!  2^n} } \vert n >.
\label{eq:xm}
\ee
Using again the Darboux-Christofell equation we can prove
the orthonormality condition of the coordinate eigenvectors:
\be
\left< x_m \vert x_{m'} \right>=
\delta_{m,m'}.
\label{eq:ortho}
\ee

The above orthonormality condition implies that the  eigenvectors
$\vert x_m >$ with $m=-J,\ldots , J $  are independent ones and they span all
the space generated by the eigenvectors $\vert n >$ with $ n=0,\ldots ,n_d-1$.
Therefore the following completeness relation holds:
\be
\sum\limits_{n=0}^{n_d-1} \vert n ><  n  \vert =
\sum\limits_{m=-J}^{J} \vert x_m ><  x_m  \vert = 1.
\label{eq:completeness}
\ee
This completeness relation implies that
$$
<n\vert n'> = \sum\limits_{m=-J}^{J} <n\vert x_m ><  x_m  \vert n'>=
\delta_{n,n'},
$$
and from the definition of eq. (\ref{eq:xm})  the orthogonality condition for
 the
Hermite polynomials is induced:
\be
\sum\limits_{m=-J}^{J}
\frac{ H_n(x_m) H_{n'}(x_m) } { {\cal N}_m } =
\delta_{n,n'} [n]! 2^n.
\label{eq:orthogonality}
\ee

In order to get a feeling about the discrete eigenvalue spectrum of
the operator $X$, let us consider the special case $n_d=4$, in which
one easily finds that $[1]=1$, $[2]=\sqrt{2}$, $[3]=1$. The relevant
equation then reads
$$ 4 x^4 -2(2+\sqrt{2}) x^2 +1 =0,$$
which has the roots
$$ \sqrt{2} x = \pm \sqrt{1+{1\over\sqrt{2}}\pm \sqrt{ \sqrt{2}+{1\over 2}}}.
$$
It is clear that the lattice occurring here is not equidistant.

Let us now consider, again in analogy to the usual harmonic oscillator case,
 the ``momentum'' operator
\be P={a-a^+\over i\sqrt{2}}.
\label{eq:P-def}
\ee
Let then $|p>$ and $p$ be the eigenvectors and the corresponding eigenvalues
of the operator $P$, satisfying
$$ P |p> = p|p>. $$
It is clear that, when $q$ is a root of unity,
$|p>$ are vectors in a $n_d$-dimensional Fock space.
Therefore, in analogy to the case of $X$, its general form is given by
\be
 |p> = \sum_{n=0}^{n_d-1} {d_n\over \sqrt{[n]!}} |n> =
         \sum_{n=0}^{n_d-1} d_n {(a^+)^n\over [n]!} |0> .
\label{eq:p}
\ee
Acting with the operator $P$ on the vectors of eq. (\ref{eq:p}) one gets,
in analogy to the $X$ case,  the following
recurrence relations for the coefficients $d_n$
$$ d_1 =           i \sqrt{2} p d_0, $$
\be
 d_{n+1} =  i \sqrt{2} p d_{n} +  [n] d_{n-1},
\label{eq:dn}
\ee
\be
d_{n_d}=0.
\label{eq:dnd}
\ee
These  recurrence relations can be satisfied only by discrete values of the
eigenvalue $p$. This shows that $q$ being a root of unity induces a
 discretization of the spectrum of the ``momentum'' operator $P$.
By comparing eq. (\ref{eq:dn}) and the definition of the
Hermite polynomials (eq. (\ref{eq:Hermite})) we conclude that:
\be
d_n(p_m)= \frac{i^{-n}2^{-n/2}}{ \sqrt{ {\cal N}_m } } H_n( -p_m),
\label{eq:rel-d-H}
\ee
where $p_m,\; m=-J,\ldots , J$ are the roots of the Hermite polynomial
of order $n_d$
$$
H_{n_d}(-p_m) =H_{n_d}(p_{-m})= 0, \quad
m=-J,\ldots , J, \quad 2J+1=n_d.
$$
We remark that these equations are identical with the ones obtained for the
eigenvalues of the $X$ operator.
Therefore the spectra of the operators $X$ and $P$ are identical and the
same q-Hermite polynomials appear in both cases.

The corresponding eigenvectors of the momentum operator are given,
in analogy to eq. (\ref{eq:xm}), by:
\be
\left\vert  p_m \right>
=
{1 \over \sqrt{{\cal N}_m }}
\sum\limits_{n=0}^{n_d-1}
{ {i^{-n} H_n \left(  -p_m \right)} \over
\sqrt{\left[ n \right]!  2^n} } \vert n >
\label{eq:pm},
\ee
and the orthogonality condition of the Hermite polynomials induces
the condition:
$$
<p_m \vert p_m'> = \delta_{m,m'}.
$$

As a result of the discretization of the ``position'' and ``momentum''
eigenvalues found above, the phase space $(x,p)$ is not the whole real
plane, as in the case of the usual harmonic oscillator, but it is a
two-dimensional lattice with non-uniformly distributed points. Similar
discretized phase spaces occur in studies of the connection between
q-deformed quantum mechanics and quantum mechanics on a lattice
\cite{Wess,LiSheng,Dimakis,Ubriaco}.

Every vector in the Hilbert space spanned by
$\vert n >,\; n=0,\ldots , n_d-1$ can be described in the ``coordinate''
representation using the basis $\vert x_m >$:
$$
\vert f> = \sum\limits_{m=-J}^{m=J}
f(x_m) \vert x_m>,
$$
or in the ``momentum'' representation using the basis $\vert p_m>$:
$$
\vert f> = \sum\limits_{m=-J}^{m=J}
{\widehat f}(p_m) \vert p_m>,
$$
where the momentum representation
${\widehat f}(p_m)$ is the ``Fourier transform'' of the
coordinate representation $f(x_m)$. This kind of Fourier transform
is given by:
\be
{\widehat f}(p_m)= \sum\limits_{m'=-J}^{J}
<p_m\vert x_{m'}> f(x_{m'}).
\label{eq:Fourier}
\ee
{}From eqs (\ref{eq:xm}) and (\ref{eq:pm}) we can define the
transition products:
\be
<p_m\vert x_m'> =
{1 \over \sqrt{ {\cal N}_m  {\cal N}_{m'}  }}
\sum\limits_{n=0}^{n_d-1}
{ {i^{n}  H_n \left(  -p_m \right)H_n \left(  x_{m'} \right)
} \over
{\left[ n \right]!  2^n} }.
\label{eq:pm-xm}
\ee
Thus the Fourier transform between the representations is completely defined.

It is worth remarking that in the present case
\be
 [X,P]= i [a,a^+] =
i\left( [N+1] - [N] \right) =
i {\cos{\pi (2N+1)\over 2 n_d} \over \cos
{\pi \over 2 n_d}}.
\label{eq:comm1}
\ee
In lowest order approximation this gives
$$
 [X,P]= i \left(1-{\pi^2\over 2 n_d^2} N(N+1)\right),
$$
which is reminiscent of the noncanonical commutation relation of ref.
\cite{Jannussis}.

All the above proposed formalism is based on the fact that the Fock space
of the q-deformed oscillator with $q$ being a root of unity is finite
dimensional. We are now going to demonstrate  that the same formalism
can be applied to other finite-dimensional oscillators as well.
Recently several generalizations of the notion of the q-deformed oscillator
have been  proposed in the literature (see \cite{BD-SUSY} for a list of
references). All these generalizations
can be   described by the deformed oscillator algebra
 generated by the
operators $\big\{ 1,a,a^+,N\big\}$ and the {\it structure
function} $\Phi (x)$, satisfying the relations \cite{Da1}:
\be
\left[ a , N \right] = a, \quad  \quad
 \left[ a^+ , N \right] = -a^+ ,
\label{eq:do1}
\ee
and
\be a^+a=\Phi(N)=[N], \qquad aa^+=\Phi(N+1)=[N+1],
\label{eq:do2}
\ee
where $\Phi(x)$ is a positive analytic function with
$\Phi(0)=0$ and $N$ is the number operator.
When
\be
\Phi(n_d)=[n_d]=0, \quad \mbox{and} \quad
a^{n_d}= \left( a^+ \right)^{n_d} =0,
\label{eq:do3}
\ee
a finite dimensional deformed oscillator algebra can be defined.
The q-deformed oscillator with q being a root of unity corresponds
 to the choice of structure function given by eq.
(\ref{eq:q-str}). Other possible choices include:

i) The parafermionic oscillator \cite{OK,BD-SUSY} with:
\be
\Phi(x) = [x] =x (n_d -x ) /(n_d -1 ).
\label{eq:sf-p}
\ee

ii) The q-deformed version \cite{FV,BD-SUSY}
 of the above parafermionic oscillator, where:
\be
\Phi(x) = [x] = \frac{ \sinh (q x) \sinh (q(n_d-x) )}
                     { \sinh (q ) \sinh (q(n_d-1) )}.
\label{eq:sf-qp}
\ee

iii) The truncated harmonic oscillator \cite{Fig}, corresponding to the
structure function:
\be
\Phi(x)= [x] = x \Theta (n_d -x),
\label{eq:tr-ho}
\ee
where $\Theta (x) $ is the step function:
\be
\Theta (x) =
\left\{
\begin{array}{ll}
0 & \mbox{if} \quad x\le 0 \\
1 & \mbox{if} \quad x>0.
\end{array}\right.
\label{eq:theta}
\ee
In all these cases the above proposed formalism can be applied by replacing
$[x]$ by the
corresponding structure function $\Phi(x)$ in all equations. Especially
in the case of the truncated harmonic oscillator \cite{Fig}, the Hermite
polynomials of eq. (\ref{eq:Hermite})
correspond to the usual Hermite polynomials, since for this oscillator
$\Phi(x) =x$ in the non-vanishing cases.

It is easy to check that in all of the above mentioned cases
the phase space is  represented by a grid with
non-equidistant points, with the exception of the parafermionic oscillator,
the phase space grid of which contains equidistant points.
In order to compare the differences of the induced inhomogeneity for the
various oscillators,
let us define the inhomogeneity ratio $\rho$
\be
\rho =
\frac{ x_{n_d} - x_{n_d-1} }{x_{k}-x_{k-1}},
\label{eq:rho}
\ee
where $k$ is the integer part of $\frac{n_d}{2}$. It is clear that
the inhomogeneity ratio is a measure of the deformation of the lattice,
by comparing the distance between two consecutive lattice points on the
boundary to the distance between two consecutive central points.
Numerical results are shown in Table 1.  The
truncated harmonic oscillator
defined by the structure function of eq. (\ref{eq:tr-ho}) induces a
deformation
larger than the one induced by the q-deformed oscillator, defined
by the structure function of eq. (\ref{eq:sf-q}).
In both of these cases the grid is more dense near the
center of the phase space lattice and more sparse near its edges.
In contrast,
the parafermionic oscillator of eq. (\ref{eq:sf-p}) corresponds to
a uniform lattice, while in the case of the q-deformed parafermionic
oscillator of eq. (\ref{eq:sf-qp})
the center of the phase space lattice is more sparse than the edges.
These facts make the  parafermionic oscillator attractive as a model
possessing an equidistant phase space grid.

As mentioned before, the operator $X$ is chosen in eq. (\ref{eq:X-def})
in analogy to the case of the usual harmonic oscillator. This choice is
quite arbitrary. More generally, one could consider an operator of the form
\be
 X ={f(N) a + g(N) a^+ \over \sqrt{2}},
\label{eq:genX}
\ee
where $f(N)$ and $g(N)$ are functions of the number operator $N$.
Following the same steps as before one sees that in this case
 the recursion
relations  of eq. (\ref{eq:cn}) for the coefficients become
$$ c_1 f(0) = \sqrt{2} x c_0, $$
$$ c_{n+1} f(n) = \sqrt{2}  x c_n - [n] c_{n-1} g(n), $$
with eq. (\ref{eq:cnd}) remaining the same. As a consequence,
 the recursion relation (\ref{eq:Hermite})
for the deformed Hermite polynomials becomes
$$ H_{n+1}(x) f(n) = 2 x H_n(x) -2 [n] H_{n-1}(x) g(n).$$
For $f(N)=1$ and $g(N)=q^{1-N}$ one obtains
$$ H_{n+1}(x)= 2 x H_n(x) -2 [n]_Q H_{n-1}(x),$$
where
\be
 [x]_Q = {Q^x -1\over Q-1}.
\label{eq:Qnum}
\ee
For $Q=q^{-2}$ this is equivalent to the recursion relation
obtained in \cite{Jeugt}, up to the factors of 2.
These factors disappear by choosing
 $$ X= f(N) a + g(N) a^+. $$
Then instead of eq. (\ref{eq:cn}) one has
$$ c_1 f(0) = x c_0, $$
$$ c_{n+1} f(n) = x c_n - [n] c_{n-1} g(n),$$
while in the place of eq. (\ref{eq:Hermite}) one obtains
$$H_{n+1}(x) f(n) = x H_n(x) -[n] H_{n-1}(x) g(n), $$
which for $f(n)=1$ and $g(N)=q^{1-N}$ reproduces the result of
ref. \cite{Jeugt} exactly, when the definition of eq. (\ref{eq:Qnum}) is used.
One therefore concludes that there are many ways of writing $X$ in
terms of $a$ and $a^+$. In all cases discretization is gotten,
although the deformed Hermite polynomials obtained differ.
One thus obtains in this way a large number of toy models.

In a similar way, one can consider
$$ P= {f(N) a - g(N) a^+ \over i \sqrt{2}}, $$
so that both $X$ and $P$ will lead to the same deformed Hermite polynomials.
In fact in this case eq. (\ref{eq:dn}) is replaced by
$$ d_1 f(0)= i\sqrt{2} p d_0, $$
$$ d_{n+1} f(n)= i\sqrt{2} p d_n + [n] d_{n-1} g(n) ,$$
and the deformed Hermite polynomials occuring are the same as in  the
case of $X$ mentioned above.

For the commutator of $X$ and $P$ one finds
$$ [X,P] = i ( f(N) g(N+1) a a^+ - f(N-1) g(N) a^+ a).$$
Using the selection of $f(N)=1$, $g(N)=q^{1-N}$, corresponding to the
case of ref. \cite{Jeugt}, one obtains
$$ [X,P]= i( [N+1]_Q -[N]_Q),$$
with $Q=q^{-2}$. This is the same result one would have obtained
using the original definitions of eqs. (\ref{eq:X-def}) and (\ref{eq:P-def})
with $a$ and $a^+$ being replaced by Q-bosons \cite{AC} $b$ and $b^+$,
 satisfying
$$ [N, b^+] = b^+, \quad [N, b] =-b,\quad  b b^+ - Q b^+ b =1.$$
This means that with the appropriate choice of $f(N)$ and $g(N)$ we
can ``convert'' q-bosons into Q-bosons, so that the result
in the commutator of $X$ and $P$ (and also in the recursion relation for the
Hermite polynomials) for $f(N)=1$, $g(N)=q^{1-N}$, and $a$, $a^+$
corresponding to the q-oscillator,
is the same as it would have been by using
$X$ and $P$ with $f(N)=1$, $g(N)=1$, and with $a$ and $a^+$ replaced by
$b$ and $b^+$, corresponding
to the Q-oscillator with $Q=q^{-2}$.
Similarly by choosing $f(N)=1$ and
$$ g(N)= { N(n_d-N) \over (n_d-1)} {(q-q^{-1}) \over (q^N -q^{-N})},$$
 the recursion relation for the deformed Hermite polynomials becomes
 the same as in the case of the parafermionic oscillator and
the spectrum is equidistant. Thus this choice of $f(N)$ and $g(N)$
 ``converts'' q-bosons into parafermions.

In conclusion, we have proved that in the case of the q-deformed oscillator
with $q$ a root of unity the ``position'' and ``momentum'' operators  have
discrete eigenvalues, which are the roots of the corresponding deformed
Hermite polynomials. The phase space is a non-equidistant grid. Similar
non-equidistant discretized phase spaces occur also in the cases of the
truncated harmonic oscillator and the q-deformed parafermionic oscillator,
while the usual parafermionic oscillator corresponds to an equidistant
phase space grid.

Using these oscillators as the basis for building
field theories defined on a uniform or non-uniform lattice seems to be
an interesting problem. Further development of the theory of q-deformed
orthogonal polynomials for $q$ being a root of unity is also needed.

%A few open questions are listed here:

%i) It should be possible to relate the $X$ and $P$ operators to some kind of
% position and its derivative. The transformation of eq. (3.3) of ref.
%\cite{Jeugt} is an idea,
%but the cases examined in ref. \cite{Jeugt} lead to q-Hermite polynomials
%different from the present ones. Other related references are
%\cite{Atakishiev,Minahan,Chang,LiSheng}.

%ii) The relation between the q-Hermite polynomials obtained here to other
%q-Hermite polynomials appearing in the literature should  be
%examined. Notice that the recursion relation of eq. (4.12) of ref.
%\cite{Jeugt}
%for $\tilde q= q^{-2}$ and $c$ chosen as in ref. \cite{Jeugt} becomes
%formally the same as eq. (\ref{eq:Hermite})
% of the present paper, but with a different definition
%of the q-number (the one of \cite{AC}).

%iii) The theory of q-deformed orthogonal polynomials for the case of
%q being a root of unity is not yet fully developed, while the case
%of q being a real number is currently the subject of investigation by
%several authors. In the case of q real the phase space should be
%described by some kind of deformed geometry, while in the finite dimensional
%deformed oscillator case the phase space is described by a non-uniformly
%distributed square grid. It also seems interesting to use the truncated
%oscillator described by the structure function of eq. (\ref{eq:tr-ho})
% as a basis for the construction of a field theory defined on a
%non-uniform lattice in the phase space.

Support from the Bundesministerium f\"ur Forschung und Technologie under
contract No 06 T\"u 736
(DB) and DGICYT, Spain (DE) is gratefully acknowledged.

\vfill\eject

\vfill\eject

\centerline{\bf Table 1}

\medskip
{\begin{center}
The inhomogeneity ratio $\rho$, defined in eq. (\ref{eq:rho}),
for various deformed oscillators.\end{center}}

\begin{center}
\begin{tabular}{|c|c|c|c|c|}
\hline
$n_d$ &
\parbox{1.0in}{\begin{center}q--deformed \\ oscillator\\
              eq. (\ref{eq:sf-q})\\
                \end{center}}
 & \parbox{1.0in}{\begin{center} truncated\\ oscillator\\
                 eq. (\ref{eq:tr-ho}) \\
                 \end{center}}
  & \parbox{1.0in}{\begin{center} q-parafermionic\\ oscillator (q=1)\\
                eq. (\ref{eq:sf-qp}) \\
                 \end{center}}
 & \parbox{1.0in}{\begin{center} parafermionic\\ oscillator\\
                eq. (\ref{eq:sf-p})\\
                 \end{center}}\\
\hline\hline
10    &  1.11 & 1.30  & 0.49 & 1  \\
\hline
15    &  1.13 & 1.45  & 0.33 & 1   \\
\hline
\end{tabular}\\
\end{center}

\end{document}